\documentclass[conference]{IEEEtran}
\IEEEoverridecommandlockouts
\usepackage{cite}
\usepackage{graphicx}
\usepackage{caption}
\usepackage{float}
\usepackage{amsmath}
\usepackage{amsthm}
\usepackage{amsfonts,dsfont}
\usepackage{amssymb,mathcomSTEv4}
\usepackage{algorithm} 
\usepackage{algpseudocode} 

\usepackage{graphicx}
\usepackage{subcaption}

\usepackage{xcolor}

\newtheorem{Theorem}{Theorem}

\newtheorem{Corollary}{Corollary}

\newtheorem{Remark}{Remark}

\def\BibTeX{{\rm B\kern-.05em{\sc i\kern-.025em b}\kern-.08em
    T\kern-.1667em\lower.7ex\hbox{E}\kern-.125emX}}
\begin{document}

\title{The Query/Hit Model for \\ Sequential Hypothesis Testing}

\author{%
  \IEEEauthorblockN{Mahshad Shariatnasab}
  \IEEEauthorblockA{School of Computing\\
  and Information Sciences \\
                    Florida International University\\
                    Miami, FL, USA\\
 mshar075@fiu.edu}

\and 
 \IEEEauthorblockN{Stefano Rini}
  \IEEEauthorblockA{Electrical and Computer\\ Engineering Department\\ 
                    National Yang Ming\\Chiao Tung University, Taiwan\\
     stefano.rini@nycu.edu.tw}
  \and
   \IEEEauthorblockN{Farhad Shirani}
  \IEEEauthorblockA{ School of Computing\\  and Information Sciences \\
                  $\!\!\!$\hspace{-.075in}  Florida International University\\
                    Miami, FL, USA\\
                 fshirani@fiu.edu}
  \and 
     \IEEEauthorblockN{S. Sitharama Iyengar}
  \IEEEauthorblockA{School of Computing\\  and Information Sciences \\
                    \hspace{.075in}Florida International University\\
                    Miami, FL, USA\\
                     iyengar@fiu.edu}

}

\maketitle

\begin{abstract}
This work introduces the Query/Hit (Q/H) learning model. The setup consists of two agents. One agent, Alice, has access to a streaming source, while the other, Bob, does not have direct access to the source. Communication between the agents is structured through sequential Q/H pairs: Bob sends a sequence of source symbols (queries), and Alice responds with the waiting time between the current time and the occurrence of each query (hits). This model is inspired by real-world scenarios where communication and privacy constraints limit real-time access to the source of information. 
The error exponent for sequential hypothesis testing under the Q/H model is characterized, 
and a querying strategy is proposed. The strategy employs mutual information neural estimators to compute the error exponent associated with each query and to select the optimal query. 
Extensive empirical evaluations on both synthetic and real-world datasets --- including mouse movement trajectories, typesetting patterns, and touch-based user interactions --- are provided to evaluate the performance of the proposed strategy in comparison with baselines, in terms of probability of error, query choice, and time-to-detection.


\end{abstract}

\section{Introduction}

Sequential hypothesis testing arises naturally in a variety of real-world scenarios where decisions must be made by an agent based on a sequence of observed data under time constraints. Applications include sequential classification, network intrusion detection, medical diagnosis, and sensor networks \cite{haghifam2021sequential,hsu2022universal,zhu2010distributed,zrnic2021asynchronous,naghshvar2013active}. 
Traditional hypothesis testing frameworks often assume direct access to the data stream, allowing the agent to gather observations until a statistically significant conclusion can be drawn. The objective in these approaches is to design a testing strategy minimizing the probability of error under time-to-detection constraints.

Many modern applications present new challenges that are not captured in the traditional framework. Specifically, privacy constraints, communication limitations, and distributed data settings create scenarios where direct access to the source of information is either infeasible or undesirable. For instance, in privacy-sensitive environments, such as medical diagnosis or user behavior analysis, the data owner (e.g., a patient or user) may not permit direct sharing of raw data, and such data collection may be further prohibited due to privacy protection laws --- such as the European Union's General Data Protection Regulation (GDPR)  and the California Consumer Privacy Act (CCPA) --- 
necessitating indirect methods for hypothesis testing. Similarly, in communication-constrained systems, such as wireless sensor networks, bandwidth limitations make continuous data transmission impractical, requiring alternative approaches that minimize communication overhead while still enabling effective inference. 

To address these challenges, we introduce a novel framework for real-time data access, classification, and hypothesis testing called the Query/Hit (Q/H) model.  In this model, two agents interact: Alice, has access to a stream of symbols generated by an ergodic process, and Bob, acts as the learner by sending queries to Alice. Bob's queries consist of sequences of source symbols, called \textit{query patterns}, and Alice responds with the waiting time until the query pattern appears in her data stream, called \textit{hit times}. Bob infers information about the source distribution indirectly through the received hit times. The model is applicable to various learning settings --- such as active learning, reinforcement learning, and change detection, among others.  

The Q/H model is driven by practical considerations such as  communication efficiency and privacy constraints. For instance, consider a scenario where a server is tasked with detecting automated agents visiting a website. 
A common approach is  to collect user interaction data, such as mouse movements, clicks, and clipboard actions (e.g., cut and paste operations), through a web socket \cite{acien2022becaptcha,ousat2024breaking}. However, privacy regulations and communication constraints make it impractical to transmit the complete collected raw interaction data back to the server. Instead, \textit{event triggers} are used to capture the occurrence and timing of specific user actions or patterns rather than transmitting detailed logs, thus enabling indirect hypothesis testing.
%
%
%
%


\paragraph{Relevant Works.}
The sequential hypothesis testing problem has been extensively studied, beginning with foundational work such as Wald's sequential probability ratio test (SPRT) \cite{wald1944cumulative}. Subsequent works have focused on adaptive testing methods, nonparametric approaches, characterization of error exponents, and the incorporation of machine learning techniques to address complex statistical scenarios 
\cite{tartakovsky2014sequential,lalitha2018social,xie2021sequential,jia2019optimal,veeravalli1994decentralized,pele2008robust}. Indirect hypothesis testing, by observing the hit times of specific patterns as in the proposed Q/H model, is novel. However, the statistical properties and information content of hit times have been studied in prior literature under various statistical models. 
In \cite{ekroot1991entropy} a closed-form expression was derived for the entropy of hit times of discrete memoryless sources. The conditional entropy of hit times of Markov trajectories conditioned on the intermediate state was characterized in \cite{kafsi2013entropy}. The asymptotic behavior of hit times, along with strong approximation theorems, were studied in \cite{kontoyiannis1998asymptotic}.
\paragraph{Contributions} We introduce the Q/H learning model, and consider
sequential hypothesis testing under the model.
We define the error exponent in terms of the ratio between the Type II error probability and the expected time to attain a desired Type I error probability. 
Using Chernoff–Stein-type arguments,  we prove that, under specific statistical assumptions, the problem of finding the optimal querying strategy can be expressed in a computable form as an optimization of
the ratio between the Kullback-Leibler (KL) divergence of the hit time under the two hypotheses divided by the expected hit time, as averaged over Bob's belief on the correct hypothesis. We call this ratio the \textit{query efficiency} ratio.
As a high-level, the query efficiency captures the tradeoff between two querying strategies: \emph{scout} vs \emph{sentinels} queries. Scout queries have low discriminative power but short expected hit times. Sentinel queries are the opposite. 

We consider both the adaptive and static setting. 
In the static setting, Bob adopts a fixed query that does not depend on Alice's query responses.  Theorem \ref{th:1} establishes the optimal query strategy for Markovian sources under the static setting. 
For the adaptive setting, we propose an algorithm, the \emph{Dynamic Scout-Sentinel Algorithm (DSSA)}, which, at any given time,  selects the query yielding the largest query efficiency given Bob's belief. 
In general real-world datasets, the query efficiency is estimated using a neural mutual information estimator. 
We conduct extensive empirical evaluations on synthetic and real-world datasets, evaluating the performance of the proposed strategies in terms of error rate, query choice, and detection time.

\begin{figure*}[!t]
    \centering    \includegraphics[width=0.75\linewidth]{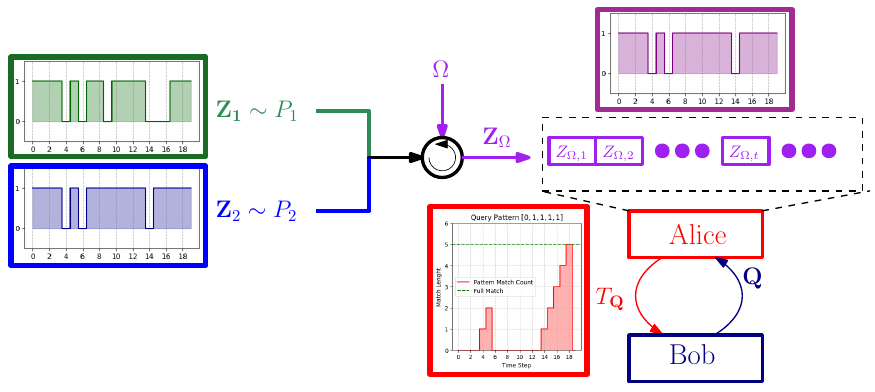}
    \caption{Overview of the Q/H  hypothesis testing problem. Two streams of symbols $\mathbf{Z}_1$ and $\mathbf{Z}_2$ are generated based on distributions $P_1$ and $P_2$, respectively. One of the hypothesis, $\mathbf{Z}_{\Omega}$ (purple plot), is observed by Alice. Bob sends a query pattern $\mathbf{Q}$. Alice observes the stream $\mathbf{Z}_{\Omega}$ until a hit, and sends back the hit time $T$.}
    \label{fig:overview}
\end{figure*}
\textit{Notation:} We use calligraphic letters, such as $\mathcal{X}$, to denote sets. The set $\{1, 2, \dots, n\}$ is represented by $[n]$. A vector $(x_1, x_2, \dots, x_n)$ is written as $x^n$, and a subvector from index $k$ to $n$ is denoted by $x_k^n$. The $i$-th element of $x^n$ is denoted by $x_i$. When the dimension is clear, we use bold-face letters, such as $\mathbf{x}$, to represent vectors. Random variables and random vectors are denoted by upper-case letters, such as $X$ and $\mathbf{X}$, respectively, while their realizations are represented by corresponding lower-case letters, such as $x$ and $\mathbf{x}$.  
\section{Problem Formulation}
\label{sec:System Model}
\subsection{The Q/H Hypothesis Testing Setup}
The setup is shown in Figure \ref{fig:overview}. We consider two sources of information represented by random sequences $\mathbf{Z}_1 = (Z_{1,1}, Z_{1,2}, \dots)$ and $\mathbf{Z}_2 = (Z_{2,1}, Z_{2,2}, \dots)$, generated according to probability measures $P_1$ and $P_2$, respectively. Both sources take values from a shared finite alphabet $\mathcal{Z}$. A switch variable $\Omega$ is randomly chosen, based on $P_{\Omega}$, from the set $\{1, 2\}$. It determines which sequence, $\mathbf{Z}_{\Omega}$, is observed by Alice. Bob's objective is to identify the observed sequence by sending \textit{query patterns} to Alice. Bob selects a pattern $\mathbf{Q} = (Q_1, Q_2, \dots, Q_m)$, where $m$ is a fixed query length, and sends it to Alice. Alice then responds with the smallest time index at which the query pattern is found in $\mathbf{Z}_{\Omega}$. The Q/H problem is completely characterized by the parameters $(P_1,P_2,m)$. 


Formally, let $\mathbf{Q}_i, i \in \mathbb{N}$, denote the sequence of $m$-length queries sent by Bob, and let $T_i, i \in \mathbb{N}$, denote the corresponding sequence of hit times returned by Alice. Bob's operation is characterized by a sequence of query functions: 
\begin{align*}
f_i: \mathbf{T}_i \times \pi_0 \mapsto \mathbf{Q}_i, i\in \mathbb{N}
\end{align*}
where  $\mathbf{T}_i=(T_j)_{j < i}$ and $\pi_0 = P_{\Omega}(1)$ represents Bob's initial belief about the value of the switch variable. The strategy is called adaptive if Bob's query pattern at time $i\in \mathbb{N}$ depends on the previous query responses $\mathbf{T}_i$, and it is called static, otherwise.

 Alice's response is determined as follows: 
 \begin{align*} T_i = \min \{ t \mid  T_{i-1} < t, Z_t^{t+m} = \mathbf{Q}_i\}, i\in \mathbb{N}
 \end{align*}
 where we have defined $T_0=0$.  
Bob then uses a sequence of decision functions: 
\begin{align*} 
d_i: \mathbf{T}_i\times \pi_0 \mapsto \widehat{\Omega}_i, i\in \mathbb{N} \end{align*} where $\widehat{\Omega}_i \in \{1, 2\}$. We denote the pair of query and decision functions used by Bob as $\mathbf{b}_i= (f_i,d_i)$. A summary of the notation is provided in Table \ref{tab:notation}.

%
%

\subsection{The Error Exponent}
 The probability of Type I and Type II errors is:
 \begin{align*} &\alpha_i(P_1,P_2,\mathbf{b}_i) = P_1(\widehat{\Omega}_i =2), 
\\&\beta_i(P_1,P_2,\mathbf{b}_i) = P_2(\widehat{\Omega}_i=1). \end{align*}
The (Type II) error exponent is defined as:
\begin{align}
\label{eq:exp}
e(P_1,P_2,,\mathbf{B})= \limsup_{i\to \infty}-\frac{\log\beta_i(P_1,P_2,\mathbf{b}_i)}{\mathbb{E}(T_i)},
\end{align}
where the expectation is with respect to $(\mathbf{Z}_1,\mathbf{Z}_2,\Omega)$ and we have defined $\mathbf{B}=(\mathbf{b}_i)_{i\in \mathbb{N}}$.

The optimal query function is defined as the one attaining the Chernoff–Stein lemma -type maximization:
\begin{align}
\label{eq:opt}
    e^*(\epsilon_1)= \sup_{\mathbf{B}\in \mathcal{B}} e(P_1,P_2,\mathbf{B}),\quad \epsilon_1\in (0,\frac{1}{2}),
\end{align}
where $\mathcal{B}$ consists of all sequences $\mathbf{B}$ such that 
$\limsup_{i\to \infty}\alpha_i(P_1,P_2,\mathbf{b}_i)\leq \epsilon_1$.

\subsection{IID Binary Example}
\label{sec:IID Binary Setting}

As an illustrative example, we consider the setting in which 
$P_1$ and $P_2$ are independent and identically distributed (IID) Bernoulli sources with parameters $p_1=1/2$ and $p_2=p$. 
Accordingly, the queries $\Qv_i$ are the $2^m$ binary sequences of length $m$ and the hit times can be obtained from ``waiting till a pattern appears'' problem through a Markov chain approach \cite{fu1996distribution}. 

An example is provided in Figure \ref{fig:overview}  for $p=\frac{4}{5}$. The green and blue plots on the left show the source realizations over twenty time-slots. 
The purple plot shows the realization of $\mathbf{Z}_{\Omega}$, where $\Omega = 2$. 
Alice observes this sequence and receives a query pattern $\mathbf{Q} = (0, 1, 1, 1)$ from Bob. 
The red plot shows the length of the partial matches of the query sequence.
That is, let us construct the Markov chain in which the state $j$ corresponds to the  $j$-th element in $\Qv$. 
Then the Markov chain transitions for state $j$ to $j+1$ at time $t$ if $Z_{t-j+1}^t$ equals $(Q_1,Q_2,\cdots,Q_j)$. 
%
%
%
When the Markov chain achieves $j=5$, the query has been hit, and the hit time is transmitted to Bob. 
One can observe the fundamental trade-off in the query choice under the Q/H model. 
Assume that $p$ is much larger that $1/2$. Then choosing the all one/all zero query yields the best discriminating power, as it has the largest difference in probability under the two hypotheses. 
However, these two queries have the longest waiting time under the alternative hypothesis. 
On the other hand, queries with smaller differences in probability have shorter waiting times under the alternative hypothesis. 
We refer to these two types of queries as \textit{sentinels} and \textit{scout} queries.
In this metaphor, one wishes to detect whether an enemy is present in their territory. A sentinel waits for a long time but is sure to spot an enemy, a scout is fast but is not sure to spot an enemy. 
With this terminology, it seems that the optimal strategy should be to adapt the current belief 
(i) Bob chooses sentinel queries if there is no strong belief on a hypothesis, while (ii) it selects sentinel queries otherwise.
As we shall argue in Section \ref{sec:Q/H over IID Hypothesis}, this is indeed the theoretically optimal strategy for the Q/H model with binary IID sources.

\begin{table}[]
    \centering
    \begin{tabular}{|c|c|}
    \hline
         switch variable & $\Omega$  \\
                 \hline 
        $i$-th query/hit &   $\Tv_i$/$\Qv_i$ \\
        \hline
        query/decision function & $f_i$, $d_i$, $\bv_i=(f_i,d_i)$ \\
        \hline
        $i$-th inter-arrival times 
         & $\Delta T_i =T_i-T_{i-1}$\\
         \hline
KL divergence of $\Delta T_i$  & $D_{n,k}$\\
         \hline
         Expected value of $\Delta T_i$ 
         & $t_{n,k}$ \\
         \hline
         Bob's belief at the $k$-th query & $\pi_k$ \\
         \hline
         query efficiency ratio & $\nu(\qv,\pi_k)$ \\
         \hline         
    \end{tabular}
    \caption{Notation Table}
    \label{tab:notation}
\end{table}

\section{Theoretical Analysis}
\label{sec:Theoretical Analysis}

In this section, we focus on static scenarios and derive the optimal error exponent and the associated sequences of queries and decision functions. In the subsequent sections, we will leverage this theoretical analysis to design querying strategies both for adaptive and static hypothesis testing. In the static setting, Bob sends a sequence of pre-determined queries $(\mathbf{q}_{i})_{i\in \mathbb{N}}$ to Alice. 
To keep the problem tractable, we further assume that the sequences $\mathbf{Z}_1,\mathbf{Z}_2$ are $\ell$-step Markov processes generated based on distributions $P_1$ and $P_2$, respectively, where $\ell\leq m$. 

First, we focus on the optimal choice of decision functions, and
we observe that, similar to direct sequential hypothesis testing, the sequence of Neyman-Pearson decision functions are optimal in the Q/H setting. 
To elaborate, given a sequence of queries $(\mathbf{q}_i)_{i\in \mathbb{N}}$, let us define the sequence of Neyman-Pearson decision functions with parameter $\lambda$ as:
\begin{align*}
    d_{\lambda,i}(\mathbf{t}_i)= 
    \begin{cases}
        1\qquad& \text{ if } \mathbf{t}_i\in \mathcal{A}_i(\lambda)\\
        2 & \text{otherwise}
    \end{cases},
\end{align*}
where we have defined $\mathbf{t}_i=(t_j)_{j<i}$, and
\begin{align*}
    \mathcal{A}_i(\lambda)=\{\mathbf{t}_i| P_1(\mathbf{T}_i=\mathbf{t}_i)\geq \lambda P_2(\mathbf{T}_i=\mathbf{t}_i)\}.
\end{align*}
Then, among all decision functions achieving equal Type I error probability, the Neyman-Pearson decision function achieves the highest error exponent $e(P_1,P_2,\mathbf{B})$ in \eqref{eq:exp}.
The proof  follows by first noting that the choice of the decision function only affects the numerator in Equation \eqref{eq:exp}, and then applying standard information theoretic arguments  (e.g., \cite[Theorem 11.7.1]{cover1999elements}).

Next, given the optimal choice of decision functions, we wish to characterize the optimal sequence of queries, achieving the highest possible error exponent. To this end, let us define the inter-arrival times $\Delta T_i =T_i-T_{i-1}, i\in \mathbb{N}$, where $T_0=0$. We note that, due to the Markovian property of the sources and the fact that $\ell\leq m$,
given $\mathbf{Q}_1$, the inter-arrival time $\Delta T_1$ is independent of all other inter-arrival times. Furthermore, given  $\mathbf{Q}_{i-1}$ and $\mathbf{Q}_{i}$, the inter-arrival time $\Delta T_{i}$ is independent of all other inter-arrival times. 
Let $\mathcal{Q}=\{\widetilde{\mathbf{q}}_1,\widetilde{\mathbf{q}}_2,\cdots,\widetilde{\mathbf{q}}_{|\mathcal{Z}|^m}\}$ denote the set of all possible query patterns of length $m$. We define:
\begin{align*}
&D_{0,n}= D_{KL}(P_1( \Delta T_1)||P_2(\Delta T_1) |\mathbf{Q}_1=\widetilde{\mathbf{q}}_n),
\\&D_{n,k}\!=\!D_{KL}(P_1(\Delta T_2)|| P_2(\Delta T_2)|\mathbf{Q}_1=\widetilde{\mathbf{q}}_n,\mathbf{Q}_2=\widetilde{\mathbf{q}}_k),\\
&t_{0,n}= \mathbb{E}(\Delta T_1|\mathbf{Q}_1=\widetilde{\mathbf{q}}_n),\\
&t_{n,k}= \mathbb{E}(\Delta T_2|\mathbf{Q}_1=\widetilde{\mathbf{q}}_n,\mathbf{Q}_2=\widetilde{\mathbf{q}}_k),
\end{align*}
where i) $D_{KL}$ denotes the Kullback-Leibler divergence, ii)
$n,k\in [|\mathcal{Z}|^m]$,  iii) $\widetilde{\mathbf{q}}_n,\widetilde{\mathbf{q}}_k\in \mathcal{Q}$, and iv) the expectation $\mathbb{E}(\Delta T_1|\mathbf{Q}_1=\widetilde{\mathbf{q}}_n)$ is defined as:
\begin{align*}
&    \mathbb{E}(\Delta T_1|\mathbf{Q}_1=\widetilde{\mathbf{q}}_n)
    =\pi_0\mathbb{E}_{P_1}(\Delta T_1|\mathbf{Q}_1=\widetilde{\mathbf{q}}_n)
    \\&+
   (1-\pi_0)\mathbb{E}_{P_2}(\Delta T_1|\mathbf{Q}_1=\widetilde{\mathbf{q}}_n).
\end{align*}

The next theorem characterizes the optimal sequence of queries and error exponent in terms of the  quantities above. The proof is provided in the Appendix \ref{sec:Proof of Theorem 1}.
\begin{Theorem}
\label{th:1}
Consider a Q/H hypothesis testing problem with Markovian sources parametrized by $(P_1,P_2,m)$.
Then, for each cycle\footnote{A cycle of length $k$ on $[n]$ is a sequence $(i_1,i_2),$ $(i_2,i_3),\cdots,(i_{k},i_1)$, where $i_j\in [n]$ and $i_j\neq i_{j'}$ for $j\neq j'$. } $\mathcal{C}$ on $[|\mathcal{Z}|^m]$, let us define the ratio:
\begin{align*}
    \mu_{\mathcal{C}}= \frac{\sum_{(i,j)\in \mathcal{C}}D_{i,j}}{\sum_{(i,j)\in \mathcal{C}}t_{i,j}}.
\end{align*}
The optimal sequence of queries, achieving maximum error exponent, is cyclic and
is given by $\mathbf{q}_i=\widetilde{\mathbf{q}}_{\overline{i}}$, where 
\begin{align*}
   & \overline{i}= i\mod |\mathcal{C}^*|,\qquad \mathcal{C}^*=\argmax_{\mathcal{C}}\mu_{\mathcal{C}}.
\end{align*} 
Furthermore, the maximum achievable error exponent is equal to $\mu_{\mathcal{C}^*}$. 
\end{Theorem}
\begin{Remark}
The proof of Theorem \ref{th:1} relies  the fact that the query patterns are not prefixes of each other.
Consequently, it can be extended naturally to Q/H scenarios with variable length queries, as long as the query patterns are restricted to a prefix-free set of sequences. 
\end{Remark}

The following corollary characterizes the optimal query when the two sources  $\mathbf{Z}_1$ and $\mathbf{Z}_2$ are IID. 
\begin{Corollary}
Consider a Q/H problem with IID sources $\mathbf{Z}_1$ and $\mathbf{Z}_2$. Then, the optimal query is fixed over time, and is given by:
\begin{align}
\label{eq:opt_IID}
    \mathbf{q}^*\!\!
    &
    =\! \argmax_{\qv \in \Qcal}   \frac{D_{KL}(P_1(\Delta T)||P_2(\Delta T)|\mathbf{q})}{\pi_0\mathbb{E}_{P_1}\!(\Delta T|\mathbf{q})\!+\!(1-\pi_0)\mathbb{E}_{P_2}\!(\Delta T|\mathbf{q})}  \\
&  =\argmax_{\qv \in \Qcal}  \ \ \nu(\qv,\pi_0),
\label{eq:QER}
\end{align}
\end{Corollary}
We refer to $\nu(\qv,\pi_0)$ as the \emph{efficiency ratio} of $\qv$ under the belief $\pi_0$. 
The term captures the core trade-off in the Q/H problem: each query contributes a certain decrease in error probability per unit of time as per the Chernoff-Stein lemma.
While the divergence term in the numerator of \eqref{eq:opt_IID} does not depend on the prior belief $\pi_0$, the numerator does. 
This implies that the efficiency ratio is a rational function of the belief. 

\section{Q/H Hypothesis Testing Algorithms}
In this section, we design algorithms for Q/H hypothesis testing in general statistical scenarios. The performance of these algorithms is evaluated through extensive empirical evaluations in the subsequent sections. 

\subsubsection*{Static Q/H for IID Sources}
As a first step, we consider binary IID sources $\mathbf{Z}_1$ and $\mathbf{Z}_2$. In this simple scenario, given the query pattern $\mathbf{q}$,
one can directly estimate the  hit time distribution, $P_i(\Delta T=t|\mathbf{q}),t\in \mathbb{N}, i\in \{1,2\}$, with arbitrary precision, and then compute the best query via Equation \eqref{eq:opt_IID}. 
In the Appendix, we provide Algorithm \ref{alg:hitting_time_pmf}, which takes a query pattern $\mathbf{q}$ of length $m$, and the Bernoulli parameter $p$ corresponding to a source $\mathbf{Z}$, and outputs the corresponding probability mass function $P(\Delta T=t|\mathbf{q}), t\in [t_{\max}]$, with $O(m t_{\max})$ computations. We then fix a threshold $\epsilon>0$, and increase $t_{\max}$ until the cumulative probability $\sum_{t\in [t_{\max}]}P(\Delta T=t_{\max}|\mathbf{q})$ exceeds $1-\epsilon$. An estimate of the hit time distribution is derived by normalization:
\begin{align*}
    \widehat{P}(\Delta T=t|\mathbf{q})=
    \frac{P(\Delta T=t|\mathbf{q})}{\sum_{t'\in [t_{\max}]}P(\Delta T=t'|\mathbf{q})}, \quad t\in [t_{\max}].
\end{align*}
Algorithm \ref{alg:non_ada_IID} takes the estimated hit time distributions for all query patterns  $\widehat{P}_1(\Delta T|\mathbf{q}),\widehat{P}_2(\Delta T|\mathbf{q}), \mathbf{q}\in \mathcal{Q}$, the initial belief probability $\pi_0=P_{\Omega}(1)$, and a threshold hyperparameter $\epsilon_t$ as input, and performs Q/H hypothesis testing. To provide an overview of the algorithm steps, Lines 2-3 use the estimate  hit time distributions to compute the KL divergence and expected hit times in Equation \eqref{eq:opt_IID}. Then, the optimal query is chosen by maximizing the ratio between the two aforementioned quantities (Line 5). Bob sends the optimal query $\mathbf{q}^*$ to Alice consecutively, and each time, he receives a hit time $\Delta T$. Bob's belief is updated using the Bayes' rule (Line 11). The process stops if Bob's belief value falls outside of the interval $(\epsilon_t,1-\epsilon_t)$, i.e. if Bob is confident that with probability more than $1-\epsilon_t$, one of the hypothesis is correct. Bob outputs the corresponding hypothesis index (Line 13). 
\begin{algorithm}[H]
\caption{Static Q/H for IID Sources}
\label{alg:non_ada_IID}
\textbf{Input:}  $\widehat{P}_1(\Delta T|\mathbf{q}),\widehat{P}_2(\Delta T|\mathbf{q}), \mathbf{q}\in \mathcal{Q}$, $\pi_0$, $\epsilon_t$ \\
\textbf{Output:} $\widehat{\Omega}$
\begin{algorithmic}[1]
\For{$\mathbf{q} \in \mathcal{Q}$}
\State $D_{\mathrm{KL}}(\widehat{P}_1 \parallel \widehat{P}_2|\mathbf{q})\gets \mathbb{E}_{\widehat{P}_1}( \log \frac{\widehat{P}_1(\Delta T = t|\mathbf{q})}{\widehat{P}_2(\Delta T = t|\mathbf{q})})$
\State $\mathbb{E}_{\widehat{P}_i}(\Delta T|\mathbf{q})\!\gets\! \sum_{t=1}^{t_{\max}} t \cdot P_i(\Delta T = t|\mathbf{q}), \quad \!\!i \!\in \!\{1, 2\}$
\EndFor
\State $\mathbf{q}^* \gets \argmax_{\mathbf{q} \in \mathcal{Q}} 
\ \ \nu(\qv,\pi_0)
$
\State $k\gets 0$
\While{$\pi_k \in (\epsilon_t, 1 - \epsilon_t)$}
\State $k\gets k+1$
\State Bob sends query pattern $\mathbf{q}^*$
\State Alice returns $\Delta T$
\State $\pi_k\gets \frac{\widehat{P}_1(\Delta T|\mathbf{q})\pi_0}{\widehat{P}_1(\Delta T|\mathbf{q})\pi_0+\widehat{P}_2(\Delta T|\mathbf{q})(1-\pi_0)}$
\EndWhile
\State \Return $\mathds{1}(\pi_k>1-\epsilon_t)+2\mathds{1}(\pi_k<\epsilon_t)$
\end{algorithmic}
\end{algorithm}
\begin{Remark}
\label{rem:Mark}
    Algorithm \ref{alg:hitting_time_pmf} in the Appendix, which computes an estimate of the probability distribution of the hit times for a given query, can be extended naturally to Markov sources. Consequently, Algorithm \ref{alg:non_ada_IID} can be generalized and be made applicable to Markov sources. These extensions are implemented in our empirical evaluations in Section \ref{sec:emp}.
\end{Remark}

\subsubsection*{Adaptive Q/H for IID Sources} 
The static algorithm can be modified to construct a greedy adaptive Q/H algorithm. To elaborate, after receiving the $k$-th query response, Bob computes the belief $\pi_k$ as in Line 11 of Algorithm \ref{alg:non_ada_IID}. He then selects the next query by optimizing as if the query will be used indefinitely, i.e., under the assumption of a static strategy. Formally, he selects:
$$\mathbf{q}^*= \argmax_{\mathbf{q} \in \mathcal{Q}}  \ \ \nu(\qv,\pi_k).
$$
The query is sent to Alice, the hit time is observed, and the belief is updated. The process repeats until the belief falls outside the interval $(\epsilon_t, 1 - \epsilon_t)$, using the same stopping criterion as in the static algorithm. This approach can be viewed as greedy, as Bob selects the best query at each step without considering future decisions. The complete process is provided in Algorithm \ref{alg:ada_IID} in the Appendix.




\subsubsection*{Adaptive Q/H for General Ergodic Sources}
\label{sec:A Q/H Algorithm for General Ergodic Sources}

In the previous section, we introduced a static algorithm for Q/H hypothesis testing over IID sources (Algorithm \ref{alg:non_ada_IID}
) and discussed its extensions to adaptive scenarios (Algorithm \ref{alg:ada_IID}, Appendix) and Markovian sources (Remark \ref{rem:Mark}). In this section, we extend Q/H hypothesis testing to general ergodic sources, where a primary challenge is the absence of an explicit estimate for the hit time probability distributions. Instead, we assume that we are provided two sequences of past observations (training data) $\mathcal{T}_i=(z_{i,1},z_{i,2},\cdots,z_{i,|\mathcal{T}_i|}), i\in \{1,2\}$, produced according to $P_i$. Our objective is to train a querying function for Bob using the training data.

\begin{algorithm}[H]
\caption{\textsc{DSSA for General Q/H Problems}}
\label{alg:general_ergodic_qh}
\textbf{Input:} $\mathcal{T}_1, \mathcal{T}_2$, $\mathcal{Q}$, $\pi_0$, $\epsilon_t$, sample size $\ell$
\\\textbf{Output:}  $\widehat{\Omega}$

\begin{algorithmic}[1]
\State // Pre-computation phase
\State $\text{MINE} \gets \text{TrainMINE}(\mathcal{T}_1, \mathcal{T}_2, \mathcal{Q})$
\State $\text{MLP}\! \gets\! \text{TrainMLP}(\mathcal{T}_1, \mathcal{T}_2, \mathcal{Q}, \ell)$ \!\!\!\Comment{Alg. \ref{alg:uni_belief_update_mlp}, Appendix}

\For{$\mathbf{q} \in \mathcal{Q}$}
    \State $D_{\text{KL}}(\widehat{P}_1 \parallel \widehat{P}_2|\mathbf{q}) \gets \text{MINE.EstimateKL}(\mathbf{q})$
    \For{$i \in \{1, 2\}$}
        \State $\mathbb{E}_{\widehat{P}_i}(\Delta T|\mathbf{q}) \gets \text{EmpMean}(\text{HitTimes}(\mathcal{T}_i, \mathbf{q}))$
    \EndFor
\EndFor
\State // Query selection and hypothesis testing phase
\State $k \gets 0$, $\pi_k \gets \pi_0$
\While{$\pi_k \in (\epsilon_t, 1 - \epsilon_t)$}
    \State $\mathbf{q}^* \gets \argmax_{\mathbf{q} \in \mathcal{Q}}  \ \ \nu(\qv,\pi_k)
    $
    \State $\Delta T \gets \text{QueryAlice}(\mathbf{q}^*)$
    \State $p \gets \text{MLP.Predict}(\Delta T, \mathbf{q}^*)$
    \State $\pi_{k+1}\! \gets\!\! \text{UpdateBelief}(\pi_k, p)$ \Comment{Alg. \ref{alg:belief_update_general}, Appendix}
    \State $k \gets k + 1$
\EndWhile
\State \Return $\widehat{\Omega} \gets \mathds{1}(\pi_k > 1-\epsilon_t) + 2\mathds{1}(\pi_k < \epsilon_t)$
\end{algorithmic}
\end{algorithm}.
We note that in Algorithm \ref{alg:non_ada_IID}, hit time distributions are used in three main steps: (i) in Line 2, the KL divergence between the distributions under the two hypotheses is computed for all possible query patterns; (ii) in Line 3, the expected hit time is calculated for each query pattern; and (iii) in Line 11, Bob’s belief is updated based on the query response. We propose alternative methods that do not require explicit hit time probabilities. The complete procedure is provided in Algorithm \ref{alg:general_ergodic_qh} and is described in the following.

\textbf{KL Divergence Estimation:}
Estimating the KL divergence between general ergodic sources has been of considerable interest in various applications. Traditional non-parametric methods include kernel density estimation (KDE) \cite{perez2008kullback} and k-nearest neighbor (k-NN) based estimators \cite{wang2009divergence}. These methods are more straightforward and computationally efficient for lower-dimensional data or settings with fewer dependencies. Adversarial methods, such as those discussed in \cite{nowozin2016f}, estimate f-divergences (a generalization of KL divergence) by training a discriminator network. 
The Mutual Information Neural Estimator (MINE) \cite{belghazi2018mine}, which leverages deep learning to estimate the KL divergence by maximizing a neural network-based lower bound on mutual information, is another effective technique. In this work, we utilize the MINE approach to estimate the KL divergence, although other methods can be considered depending on the statistical complexity and size of the training data. MINE leverages a neural network to learn a lower-bound estimate of MI, based on the Donsker-Varadhan (DV) representation of the KL divergence. Loosely speaking, the KL divergence  $D(P\parallel Q)$ can be expressed as:
\[
D_{KL}(P\parallel Q) =\sup_{\theta}  \mathbb{E}_P[M_{\theta}(Z)] - \log\mathbb{E}_Q[e^{M_\theta(Z)}],
\]
where $M_{\theta}:\mathcal{Z}\to \mathbb{R}$, and $\theta$ is a paramterization of (a rich enough) function space. MINE finds the optimizing function $M_{\theta}$ by training a neural network, thus finding an estimate of the KL divergence.
\\\textbf{Expected hit time Estimation:} The expected hit times can be estimated via the empirical average of the hit times in the training set. The estimate is accurate as long as the size of the training set is significantly larger than the total number of query patterns $|\mathcal{Z}|^m$. 
\\\textbf{Belief Update Estimation.} To update the beliefs, one can train a regression mechanism, to estimate $P(\Omega=i|\Delta T=t, \mathbf{q})$. Due to the highly non-linear nature of  $P(\Omega=i|\Delta T=t, \mathbf{q})$ we train a multi-layer perceptron (MLP).  The process of training and application of the MLP is described in the Appendix.

\begin{figure*}[!t]
    \centering
    \includegraphics[width=0.85\linewidth]{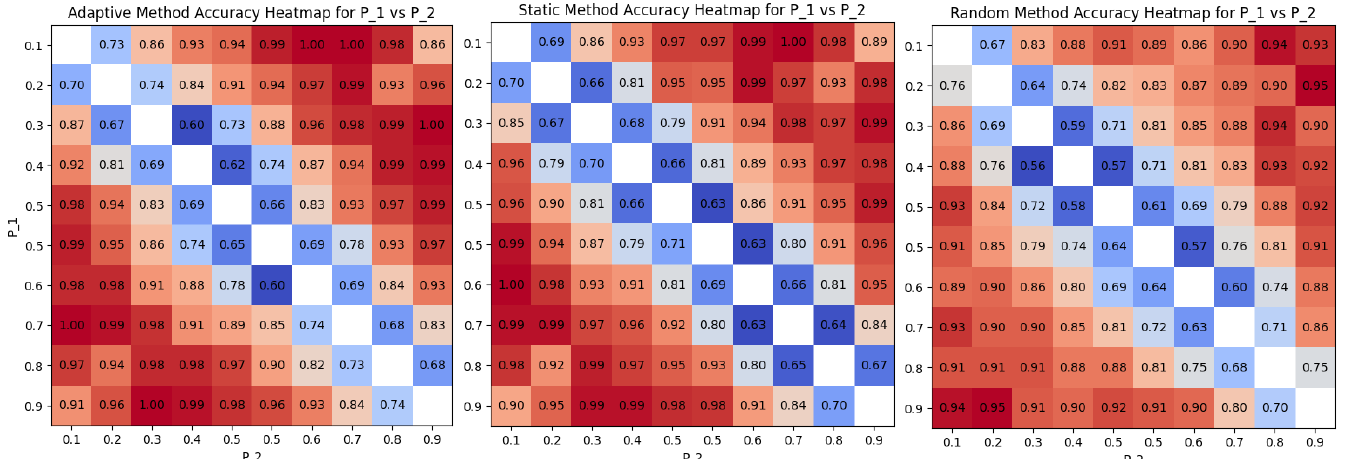}
    \caption{Accuracy heatmaps:  heuristic adaptive, optimal static, and Random Choice algorithms in IID settings.}
    \label{fig:IID_Heatmap}
\end{figure*}
\begin{figure*}[!t]
    \centering
    \includegraphics[width=0.85\linewidth]{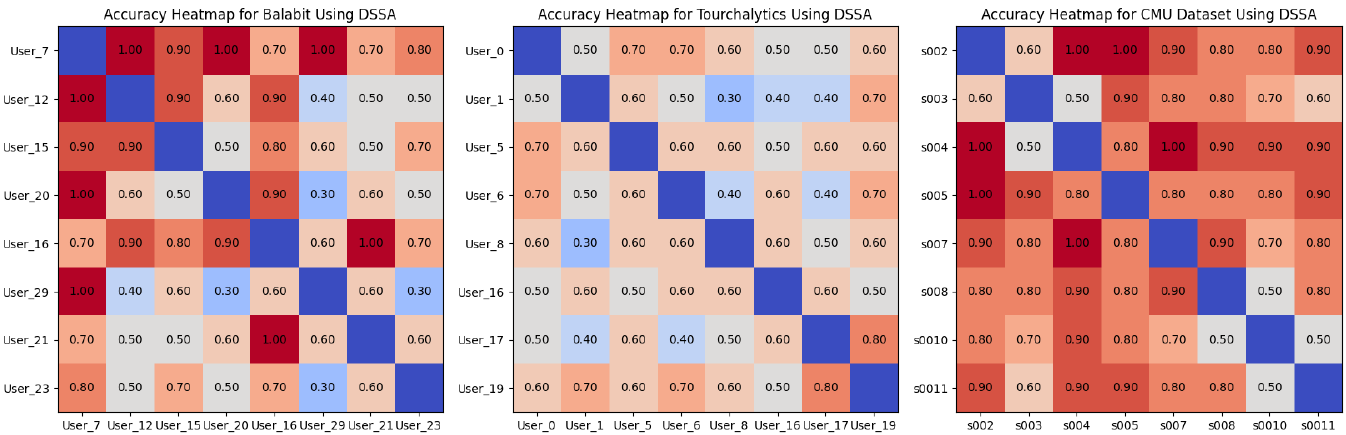}
    \caption{Heatmaps showing prediction accuracy of DSSA over the Balabit, Touchalytics, and CMU datasets.}
    \label{fig:heatmap_datasets}
\end{figure*}

\section{Experimental Evaluation}
\label{sec:emp}
We conduct extensive empirical evaluations on both synthetic and real-world datasets to gain insights into various aspects of the Query/Hit (Q/H) model and to assess the performance of the DSSA. 
In particular, we aim to address the following research questions:

\noindent \textbf{Research Question 1.} How do the efficiency ratio and the choice of the optimal query vary as a function of Bob's belief?

\noindent \textbf{Research Question 2.} How does the memory structure of the source affect the query choice and time-to-detection? Specifically, how does the performance differ when the source is  IID versus Markovian?

\noindent \textbf{Research Question 3.} How does the performance, in terms of detection accuracy and time-to-detection, compare between adaptive and static algorithms?

\noindent \textbf{Research Question 4.} How does the DSSA algorithm perform, in terms of accuracy and efficiency, when applied to real-world datasets?

\noindent \textbf{Overview of Datasets:} To answer these, we perform experiments over i) IID binary sources, ii) binary first-order Markov sources, iii) the Balabit mouse movement dataset \cite{fulop2016balabit},
iv) the Tourchalytics touch interaction dataset \cite{killourhy2009comparing}, and iv) the CMU password dataset \cite{killourhy2009comparing}. 

\begin{figure*}[!t]
    \centering
    \includegraphics[width=0.9\linewidth]{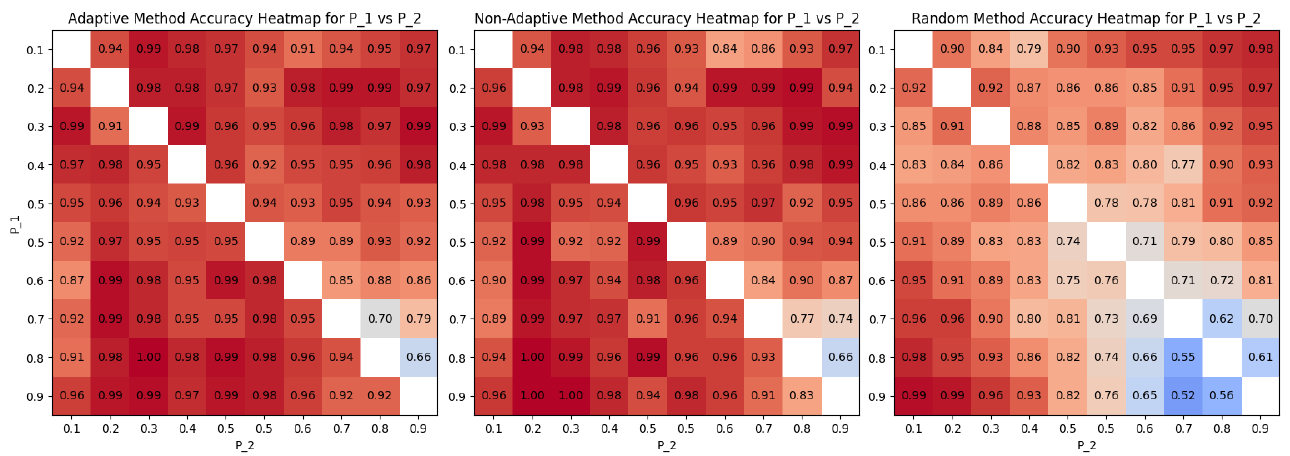}
    \caption{Heuristic adaptive, optimal static, and Random Choice algorithms in Markov settings.}
    \label{fig:heatmap_markov}
\end{figure*}

\noindent \textbf{Overview of Hypothesis Tesitng Algorithms.} To provide a baseline for comparison, we perform experiments using a hypothesis testing algorithm with random queries at each time step. To elaborate, after receiving each query response, this baseline algorithm, referred to as the \textit{Random Choice algorithm}, updates its belief using an MLP, similar to the operation of DSSA, and then sends a randomly and uniformly chosen query pattern to Alice, in contrast to the MINE-based approach of DSSA. This baseline is justified as follows.
DSSA assumes that the inter-arrivals time are IID. This assumption does not hold in general and is only asymptotically true for ergodic sequences \cite{kontoyiannis1998asymptotic}. 
In real-world datasets, the waiting times could be arbitrarily correlated in time. 
Showing the DSSA algorithm outperforms the Random Choice algorithm implies that the efficiency ratio is indeed a robust query selection policy, despite the existance of such correlations.

A complete description of the aforementioned datasets, and the MINE and MLP model architectures used in the implementation of the DSSA and Random Choice algorithms is provided in the Appendix. 

\subsection{Experiments over Synthetic Datasets}

\subsubsection{Q/H over IID Hypothesis}
\label{sec:Q/H over IID Hypothesis}

In Section \ref{sec:IID Binary Setting}, we characterized the error exponent and the sequence of optimal query functions for the static Q/H problem with IID sources. We further introduced (an adaptive) Algorithm \ref{alg:ada_IID} in the Appendix. To evaluate the effectiveness of the query choice on the accuracy performance, we have experimentally evaluated the accuracy of the optimal static, heuristic adaptive, and Random Choice baseline algorithms in Figure \ref{fig:IID_Heatmap}. We have limited the length of the sequence observed by Alice to 20 bits, and the number of queries sent by Bob to 10, and performed 400 runs of the hypothesis test for each pair of $P(Z_1=1),P(Z_2=1)\in \{\frac{i}{10}\}_{i\in [9]}$ with $m=3$. It can be observed that the adaptive algorithm slightly improves upon the static, and both outperform the random choice baseline as expected. We have provided extensive empirical evaluations of various aspects of these algorithms in the appendix.

\subsubsection{Q/H over Binary Markov Sources}
We expect the proposed adaptive and static algorithms to outperform Random Choice significantly when there are dominating patterns in the data, which can then be detected using the efficiency ratio. To demonstrate this, we consider Markov sources with distribution $$P_{Z_{i,j}|Z_{i,j-1}}(z|z')=\frac{1}{2}+(-1)^{\mathds{1}(z=z')}(0.4+0.1p),$$
where $i\in \{1,2\}$, $j\in \mathbb{N}$, $z,z'\in \{0,1\}$, and $p\in \{\frac{i}{10}\}_{i\in [9]}$. That is, the probability of consecutively equal outcomes is between 0.9 and 1 depending on  $p$. Figure \ref{fig:heatmap_markov} shows that in this scenario the proposed adaptive and static algorithms significantly outperform the Random Choice algorithm as expected. Here we have limited Alice's observations to 10 samples and $m=3$.

\subsection{Experiments over Real-world Datasets}
We have applied the DSSA and random choice algorithms on three real-world datasets, namely the Balabit, Touchalytics, and CMU password datasets. 
The dataset description, preprocessing steps, and model architectures used in our simulations are provided in the appendix. For each dataset, we have randomly chosen a subset of eight users. In each experiment, we take a pair of users from this subset and conduct a total of ten Q/H hypothesis tests on their corresponding sequences. Figure \ref{fig:heatmap_datasets} shows the resulting accuracy heatmap for the three aforementioned datasets, for DSSA with pattern length $m=2$. It can be observed that the DSSA algorithm achieves average accuracy of $68\%$ on the Balabit dataset, $56\%$ over Touchalytics dataset,
and $80\%$ on the CMU password dataset. In contrast, Figure \ref{fig:heatmap_datasets_random} in the Appendix shows the accuracy heatmap for the Random Choice baseline with pattern length $m=2$. It can be observed that the Random Choice algorithm achieves average accuracy of close to $50\%$ for all three datasets. Consequently, the performance of the DSSA algorithm is significantly better than Random Choice.

\section{Conclusion}
%
We have introduced the Q/H indirect learning model.
%
In this setting, the Chernoff-Stein lemma provides an intuitive framework for query selection based on the \textit{query efficiency ratio}. This ratio is defined as the KL divergence between the inter-arrival times of query hits under the null and alternative hypotheses, divided by the average inter-arrival time based on the current belief.
Subsequently, we have proposed the DSSA as an effective approach for sequential hypothesis testing within the Q/H model. We have presented both theoretical guarantees and extensive empirical evaluations, using both synthetic and real-world datasets, to evaluate the DSSA’s performance. The Q/H learning model presents a novel learning framework --- with applications in classification, regression, and hypothesis testing --- which captures privacy and communication constraints inherent in various emerging applications. Classification and regression under the Q/H model is an interesting avenue of future research.

\bibliographystyle{apalike}
\bibliography{References}

\appendix

\section{Proof of Theorem \ref{th:1}}
\label{sec:Proof of Theorem 1}

\begin{proof}
Fix a sequence of queries $\mathbf{q}=(\mathbf{q}_i)_{i\in \mathbb{N}}$, and let $\mathbf{T}_i= (T_1,T_2,\cdots,T_i)$ be the received query responses up to time $i$ and recall that $\Delta T_i= T_i-T_{i-1}, i\geq 1$ are the inter-arrival times, where  $T_0=0$. We have:
\begin{align*}
    &P_j(\Delta \mathbf{T}_i=\Delta \mathbf{t}_i|\mathbf{q})=
    \prod_{i'\in [i]}\!\! P_j(\Delta T_{i'}=\Delta t_{i'}|\mathbf{q}_{i'-1},\mathbf{q}_{i'}),
\end{align*}
where $j\in \{1,2\}$ and $\Delta \mathbf{t}_i\geq 0$. As a result,
    \begin{align*}      
&\log\frac{P_1(\Delta\mathbf{T}_i=\Delta \mathbf{t}_i|\mathbf{q})}{P_2(\Delta\mathbf{T}_i=\Delta \mathbf{t}_i|\mathbf{q})}
\\&= \sum_{i'\in [i]} \log{\frac{P_1(\Delta T_{i'}=\Delta t_{i'}|\mathbf{q}_{i'-1},\mathbf{q}_{i'})}{P_2(\Delta T_{i'}=\Delta t_{i'}|\mathbf{q}_{i'-1},\mathbf{q}_{i'})}}.
    \end{align*}
 Then, using standard concentration of measure arguments (e.g., using Chebychev's inequality), there exists a sequence $\epsilon_{i},i\in \mathbb{N}$ of positive numbers such that $\epsilon_i\to 0$ as $i\to \infty$, and
\begin{align*}
&P_1(\big| \log\frac{P_1(\Delta\mathbf{T}_i=\Delta \mathbf{T}'_i|\mathbf{q})}{P_2(\Delta\mathbf{T}_i
=\Delta \mathbf{T}'_i|\mathbf{q})}
- \overline{D}'_i|>\epsilon_i)\to 0,
\end{align*}
as $i\to \infty$, where the outer probability is with respect to $\mathbf{T}'_i$, and we have defined:
\begin{align*}
&\overline{D}'_i=\sum_{i'\in [i]} D'_{i'-1,i'},\\
&D'_{i'\!-\!1,i'}\!=\!D_{KL}(P_1(\Delta T_{i'}|\mathbf{q}_{i'-1},\mathbf{q}_{i'})|| P_2(\Delta T_{i'}|\mathbf{q}_{i'-1},\mathbf{q}_{i'})),
\\&D'_{0,1}= D_{KL}(P_1( \Delta T_1|\mathbf{q}_1)||P_2(\Delta T_1|\mathbf{q}_1)).
\end{align*}
Furthermore, 
\begin{align*}
&P_2(\big| \log\frac{P_1(\Delta\mathbf{T}_i=\Delta \mathbf{T}'_i|\mathbf{q})}{P_2(\Delta\mathbf{T}_i
=\Delta \mathbf{T}'_i|\mathbf{q})}
- \overline{D}'_i|<\epsilon_i)\leq 2^{-\overline{D}'_i-\epsilon_i}.
\end{align*}
Consequently, by taking $\lambda= \overline{D}'_i-\epsilon_i$ in the Neyman-Pearson decision function, we get:
\begin{align*}
   -\log\beta(P_1,P_2,m)\geq \overline{D}'_i-\epsilon_i.
\end{align*}
Furthermore, from the converse argument in the proof of the Chernoff-Stein lemma used in direct hypothesis testing (e.g., \cite[Theorem 11.8.3]{cover1999elements}), we get 
\begin{align*}
   -\log\beta(P_1,P_2,m)\leq \overline{D}'_i+\epsilon_i.
\end{align*}
On the other hand, we have:
\begin{align*}
    \mathbb{E}({T}_i|\mathbf{q}) = \sum_{i'\in [i]} \mathbb{E}(\Delta T_{i'}|\mathbf{q}_{i'-1},\mathbf{q}_{i'})= \sum_{i'\in [i]}t'_{i'-1,i'},
\end{align*}
where we have defined $t'_{i'-1,i'}=\mathbb{E}(\Delta T_{i'}|\mathbf{q}_{i'-1},\mathbf{q}_{i'})$ for $i'>1$ and $t'_{0,1}=\mathbb{E}(T_1|\mathbf{q}_1)$. So, we have:
\begin{align}
e(P_1,P_2,\mathbf{B})= \limsup_{i\to \infty} \frac{\sum_{i'\in [i]} D'_{i'-1,i'}}{\sum_{i'\in [i]}t'_{i'-1,i'}}.
    \label{eq:th1:1}
\end{align}
Let $\mathsf{C}$ denote the set of all possible cycles on $[|\mathcal{Z}|^m]$. Then, the numerator and denominator in Equation \eqref{eq:th1:1} can be decomposed as:
\begin{align*}
 \frac{\sum_{i'\in [i]} D'_{i'-1,i'}}{\sum_{i'\in [i]}t'_{i'-1,i'}}
 = 
 \frac{D'_{0,1}+ n_k\sum_{\mathcal{C}_k\in \mathsf{C}}\sum_{(i,j)\in \mathcal{C}} D_{i,j}}{t'_{0,1}+n_k\sum_{\mathcal{C}_k\in \mathsf{C}}\sum_{(i,j)\in \mathcal{C}} t_{i,j}},
\end{align*}
which asymptotically upper-bounded by $\mu_{\mathcal{C}^*}$. This concludes the proof. 
\end{proof}

\section{Computing the hit time PMF in the IID Setting}
\label{app:Computing the hit time PMF in the IID Setting}

We use a dynamic programming technique to compute an estimate of the hit time distribution for each hypothesis. 
The algorithm computes the probability mass function (PMF) for the first occurrence (hit time) of a specific binary pattern $\mathbf{q} = (q_1, q_2, \dots, q_m)$ in a sequence of independent Bernoulli trials, with Bernoulli parameter $p$. The complete process is provided in Algorithms \ref{alg:hitting_time_pmf} and \ref{alg:failure_function}, and 
an overview of its operation is provided in the following.
\begin{itemize}
    \item \textbf{Failure Function (Algorithm \ref{alg:failure_function}):} 
    The Q/H process can be viewed as a Markov chain, where the state at time $t$, denoted by $S_t$, represents the number of symbols in the pattern $\mathbf{q}$ that have been matched by that time. Specifically, if $S_t = k$, this implies that $Z_{t-k+1}^t = q_1^k$, i.e., the last $k$ symbols in the observed sequence correspond to the first $k$ symbols of the pattern $\mathbf{q}$.
If, at time $t+1$, the next Bernoulli trial outcome $Z_{t+1}$ matches the next symbol of the pattern, i.e., $Z_{t+1} = q_{k+1}$, the state transitions to $S_{t+1} = k + 1$. However, in the event of a mismatch, i.e., when $Z_{t+1} \neq q_{k+1}$, the state updates to $S_{t+1} = k'$, where $k'$ is the length of the longest proper prefix of the pattern that is also a suffix of the current partial match. The \textit{failure function} is an $m$-length vector that stores the value of $k'$ for each $k\in [m]$. The failure function is commonly used in the Knuth-Morris-Pratt (KMP) string matching algorithm \cite{knuth1977fast}. We compute the failure function using Algorithm \ref{alg:failure_function}. The computation is used in Line 1 of Algorithm \ref{alg:hitting_time_pmf} to construct  the vector $\text{failure}[1 \dots m]$.
    \item \textbf{Dynamic Programming (DP) Table:} The DP table $\text{dp}[k]$ in Algorithm \ref{alg:hitting_time_pmf} represents the probability that the first $k$ symbols of the pattern have been matched at a given time step, i.e., at time $t$ the table holds $P_{S_t}(k), k\in [m]$. Initially, $\text{dp}[0] = 1$ (since no symbols have been matched at the start), and $\text{dp}[k] = 0$ for all $k > 0$. The table is updated at each time step based on whether the outcome of the current Bernoulli trial matches the next symbol of the pattern (Lines 22 and 26).

\begin{algorithm}[H]
\caption{\textsc{HittingTimePMF}}
\label{alg:hitting_time_pmf}
\textbf{Inputs:} Pattern $\mathbf{q} = (q_1, q_2, \dots, q_m)$, Bernoulli parameter $p$, maximum time $t_{\text{max}}$ \\
\textbf{Outputs:}  $\text{PMF}[1 \dots t_{\text{max}}]$
\begin{algorithmic}[1]
\State $\text{failure}[1 \dots m] \gets \textsc{BuildFailureFunction}(\mathbf{q})$
\State $\text{dp}[0 \dots m-1] \gets 0$, \quad $\text{dp}[0] \gets 1$
\State $\text{pmf}[1 \dots t_{\text{max}}] \gets 0$
\For{$t = 1$ \textbf{to} $t_{\text{max}}$}
    \State $\text{new\_dp}[0 \dots m-1] \gets 0$
    \For{$k = 0$ \textbf{to} $m-1$}
        \If{$\text{dp}[k] = 0$}
            \State \textbf{continue}
        \EndIf
        \For{$s \in \{0, 1\}$}
            \State $q \gets \begin{cases} p & \text{if } s = 1 \\ 1 - p & \text{if } s = 0 \end{cases}$
            \State $j \gets k$
            \While{$j > 0$ \textbf{and} $s \neq q_{j+1}$}
                \State $j \gets \text{failure}[j]$
            \EndWhile
            \If{$s = q_{j+1}$}
                \State $j \gets j + 1$
            \EndIf
            \If{$j = m$}
                \State $\text{pmf}[t] \gets \text{pmf}[t] + \text{dp}[k] \times q$
            \Else
                \State $\text{new\_dp}[j] \gets \text{new\_dp}[j] + \text{dp}[k] \times q$
            \EndIf
        \EndFor
    \EndFor
    \State $\text{dp} \gets \text{new\_dp}$
\EndFor
\State \Return $\text{pmf}$
\end{algorithmic}
\end{algorithm}

    \item \textbf{Probability Updates:} At each time step $t$, for each state $S_t=k$, the algorithm computes the probability of transitioning to a new state depending on the outcome of the Bernoulli trial $Z_t$. Specifically, if the next trial outcome matches the next symbol of the pattern, i.e., $Z_t = q_{k+1}$, then the state transitions to $k+1$. Otherwise, the failure function $\textsc{BuildFailureFunction}$ is invoked to determine the state transision. Formally, the probability of a hit at time $t$ is:
\begin{align*}
&P_{\Delta T}(t) = \sum_{k=0}^{m-1} P_{Z_t|S_{t-1}}(q_m \mid  k) \cdot P_{S_{t-1}}(k),
\end{align*}
So, if $Z_t = q_m$, then the PMF is updated by incrementing $\text{pmf}[t]$ by the corresponding DP entry $\text{dp}[k] \cdot P(s)$ (Line 20). Otherwise, the dynamic probability table is updated (Lines 22 and 26).

\begin{algorithm}[H]
\caption{\textsc{BuildFailureFunction}}
\label{alg:failure_function}
\textbf{Inputs:} Pattern $\mathbf{q} = (q_1, q_2, \dots, q_m)$ \\
\textbf{Outputs:} $\text{failure}[1 \dots m]$
\begin{algorithmic}[1]
\State $\text{failure}[1] \gets 0$
\State $k \gets 0$
\For{$i = 2$ \textbf{to} $m$}
    \While{$k > 0$ \textbf{and} $q_{k+1} \neq q_i$}
        \State $k \gets \text{failure}[k]$
    \EndWhile
    \If{$q_{k+1} = q_i$}
        \State $k \gets k + 1$
    \EndIf
    \State $\text{failure}[i] \gets k$
\EndFor
\State \Return $\text{failure}$
\end{algorithmic}
\end{algorithm}

\item \textbf{Stopping Criterion.} The process continues until the maximum time $t_{\text{max}}$ is reached. The algorithm returns $\text{pmf}[1 \dots t_{\text{max}}]$, where $\text{pmf}[t]$ represents the probability that the pattern $\mathbf{q}$ first appears at time $t$. We increase $t_{\max}$ until the cumulative probability of $\text{pmf}[1 \dots t_{\text{max}}]$ reaches $1 - \epsilon$, for some $\epsilon>0$, and then normalize the vector so that the values add to one. 
\end{itemize}
The computational complexity of the above algorithm is $O(mt_{\max})$.

\section{Adaptive Q/H Algorithm for IID Sources}
\begin{algorithm}[H]
\caption{Adaptive Q/H for IID Sources}
\label{alg:ada_IID}
\textbf{Input:}  $\widehat{P}_1(\Delta T|\mathbf{q}),\widehat{P}_2(\Delta T|\mathbf{q}), \mathbf{q}\in \mathcal{Q}$, $\pi_0$, $\epsilon_t$ \\
\textbf{Output:} $\widehat{\Omega}$
\begin{algorithmic}[1]
\For{$\mathbf{q} \in \mathcal{Q}$}
\State $D_{\mathrm{KL}}(\widehat{P}_1 \parallel \widehat{P}_2|\mathbf{q})\gets \mathbb{E}_{\widehat{P}_1}( \log \frac{\widehat{P}_1(\Delta T = t|\mathbf{q})}{\widehat{P}_2(\Delta T = t|\mathbf{q})})$
\State $\mathbb{E}_{\widehat{P}_i}(\Delta T|\mathbf{q})\!\gets\! \sum_{t=1}^{t_{\max}} t \cdot P_i(\Delta T = t|\mathbf{q}), \quad \!\!i \!\in \!\{1, 2\}$
\EndFor
\State $k\gets 0$
\While{$\pi_k \in (\epsilon_t, 1 - \epsilon_t)$}
\State $\mathbf{q}^* \gets \argmax_{\mathbf{q} \in \mathcal{Q}} 
\ \ \nu(\qv,\pi_k)
$
\State $k\gets k+1$
\State Bob sends query pattern $\mathbf{q}^*$
\State Alice returns $\Delta T$
\State $\pi_k\gets \frac{\widehat{P}_1(\Delta T|\mathbf{q})\pi_0}{\widehat{P}_1(\Delta T|\mathbf{q})\pi_0+\widehat{P}_2(\Delta T|\mathbf{q})(1-\pi_0)}$
\EndWhile
\State \Return $\mathds{1}(\pi_k>1-\epsilon_t)+2\mathds{1}(\pi_k<\epsilon_t)$
\end{algorithmic}
\end{algorithm}

\section{Computing Updated Belief Using the MLP}
\label{sec:Computing Updated Belief Using the MLP}

To prepare a training set for the MLP, for each query pattern $\mathbf{q}$ and hypothesis $i \in \{1, 2\}$, and given sequences $\mathcal{T}_i = (z_{i,1}, \ldots, z_{i,|\mathcal{T}_i|})$ drawn from distributions $P_i$. We generate a dataset $\mathcal{D}$ of size $\ell\in\mathbb{N}$ as follows: i) Randomly select $\ell$ starting points within each $\mathcal{T}_i$. ii) Compute hit times $\Delta T_i$ for $\mathbf{q}$ from these points.iii) Label each sample with its corresponding hypothesis $i$. The resulting dataset $\mathcal{D}$ consists of triples $(\Delta T_i, \mathbf{q}, i)$.

We train an MLP on $\mathcal{D}$ using cross-entropy loss (Algorithm \ref{alg:uni_belief_update_mlp}). It estimates $P_{\theta}(\Omega = i \mid \Delta T, \mathbf{q})$, assuming an initial uniform belief. The loss function is:
$$
\mathcal{L}(\theta) = \frac{1}{|\mathcal{D}|} \sum_{(\Delta T, \mathbf{q}, y) \in \mathcal{D}} \text{CE}(y, P_{\theta}(\Omega = 1 \mid \Delta T, \mathbf{q}))
$$
where CE denotes the binary cross-entropy loss.

\begin{algorithm}[H]
\caption{\textsc{UniformBeliefMLP}}
\label{alg:uni_belief_update_mlp}
\textbf{Input:} $\mathcal{T}_1, \mathcal{T}_2$, $\mathcal{Q}$
\\\textbf{Output:} Trained MLP $P_{\theta}(\Omega = i \mid \Delta T, \mathbf{q})$
\begin{algorithmic}[1]
\State $\mathcal{D} \gets \emptyset$
\For{$i \in \{1, 2\}$}
    \For{$\mathbf{q} \in \mathcal{Q}$}
        \For{$j = 1$ to $|\mathcal{T}_i|$}
            \State $\text{START} \sim \text{RAND}(\mathcal{T}_i)$
            \State $\Delta T_i \gets \text{ComputeHitTime}(\mathcal{T}_i, \text{START}, \mathbf{q})$
            \State $\mathcal{D} \gets \mathcal{D} \cup \{(\Delta T_i, \mathbf{q}, i)\}$
        \EndFor
    \EndFor
\EndFor
\State $\theta^* \gets \argmin_{\theta} \mathcal{L}_{\text{CE}}(\theta; \mathcal{D})$
\State \Return $\theta^*$
\end{algorithmic}
\end{algorithm}

For a general, non-uniform initial belief $P(\Omega = 1)=s$, we derive the posterior belief using Bayes' rule and the previously trained MLP (Algorithm \ref{alg:belief_update_general}).

Let us assume that we have trained an MLP which takes $\mathbf{q}$ and $\Delta T$ as input, an outputs the updated belief $P_{eq}(\Omega=1|\Delta T,\mathbf{q})$ assuming uniform initial belief. Then, 
\begin{align*}
   & P_{eq}(\Omega=1|\Delta T,\mathbf{q})
   \\&= 
    \frac{\frac{1}{2}P(\Delta T|\Omega=1,\mathbf{q})}{\frac{1}{2}P(\Delta T|\Omega=1,\mathbf{q})+\frac{1}{2}P(\Delta T|\Omega=2,\mathbf{q})}
    \\&\Rightarrow \frac{P(\Delta T|\Omega=2,\mathbf{q})}{P(\Delta T|\Omega=1,\mathbf{q})}= \frac{1}{P_{eq}(\Omega=1|\Delta T,\mathbf{q})}-1.
\end{align*}
On the other hand, if the initial belief is not uniform and is equal to $s\in [0,1]$, then:
\begin{align*}
   & P(\Omega=1|\Delta T,\mathbf{q})
   \\&= 
    \frac{sP(\Delta T|\Omega=1,\mathbf{q})}{sP(\Delta T|\Omega=1,\mathbf{q})+(1-s)P(\Delta T|\Omega=2,\mathbf{q})}
    \\&\Rightarrow P(\Omega=1|\Delta T,\mathbf{q})= \frac{s}{s+(1-s)(\frac{1}{P_{eq}(\Omega=1|\Delta T,\mathbf{q})}-1)}.
\end{align*}

\begin{algorithm}[H]
\caption{\textsc{GeneralBeliefMLP}}
\label{alg:belief_update_general}
\textbf{Input:} $\Delta T$, $\mathbf{q}$, $P(\Omega = 1)$,$\mathcal{T}_1, \mathcal{T}_2$, $\mathcal{Q}$
\\\textbf{Output:} $P(\Omega = 1 \mid \Delta T, \mathbf{q})$
\begin{algorithmic}[1]
\State $\theta^* \gets \textsc{UniformBeliefMLP}(\mathcal{T}_1,\mathcal{T}_2,\mathcal{Q})$
\State $p\gets P_{\theta^*}(\Omega = i \mid \Delta T, \mathbf{q})$ 
\State $r \gets \frac{1}{p} - 1$
\State $P(\Omega = 1 \mid \Delta T, \mathbf{q}) \gets \frac{P(\Omega = 1)}{P(\Omega = 1) + (1 - P(\Omega = 1)) r}$
\State \Return $P(\Omega = 1 \mid \Delta T, \mathbf{q})$
\end{algorithmic}
\end{algorithm}

\begin{figure*}[!t]
    \centering
    \includegraphics[width=\linewidth]{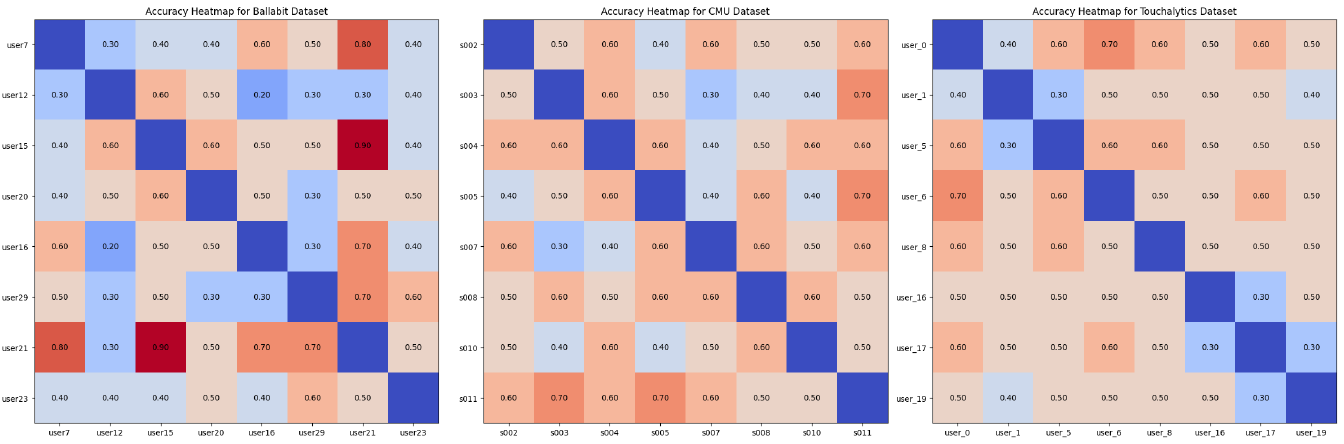}
    \caption{Heatmaps showing prediction accuracy of Random Choice over the Balabit, Touchalytics, and CMU datasets.}
    \label{fig:heatmap_datasets_random}
\end{figure*}

\section{Detailed Experimental Results}

\subsection{Dataset Description}

\smallskip
\noindent \textbf{IID Dataset Description:}
This is the simplest dataset we consider: the source symbols are generated iid Bernoulli-distributed with parameter $p_1$ and $p_2$ for hypothesis $1$ and $2$, respectively.

\smallskip
\noindent\textbf{Markov Dataset Description:}


This dataset represents a variation of the IID setting, where the IID sequence serves as the innovation sequence for a binary symmetric Markov chain. 
The result is a two-state Markov chain, where the state transition probabilities vary under each hypothesis: under hypothesis 1, the state changes with probability \( p_1 \) and remains the same with probability \( 1 - p_1 \); under hypothesis 2, the state changes with probability \( p_2 \) and remains the same with probability \( 1 - p_2 \). Formally, 
\begin{align*}
    P_{Z_{i,j}|Z_{i,j-1}}(z|z')=(1-p_i)\mathds{1}(z=z')+p_i\mathds{1}(z\neq z'),
\end{align*}
where $i\in \{1,2\}$, $j\in \mathbb{N}$, and $z,z'\in \{0,1\}$.

\smallskip
\noindent \textbf{Balabit Dataset:} The Balabit dataset was published in August 2016. It was developed for a mouse dynamics data science competition. The goal of the challenge was to protect a group of users from unauthorized access to their accounts by learning the unique characteristics of their mouse usage patterns. The dataset is organized into two folders: one for training files and another for test files. Data were collected from ten users who operated over remote desktop clients connected to remote servers. The original user identifiers included in the dataset are 7, 9, 12, 15, 16, 20, 21, 23, 29, and 35. We choose a subset of eight users, as shown in Figure \ref{fig:heatmap_datasets} for our experiments.
Each record contains the following components: 
\textbf{(i)} \textit{x} and \textit{y} coordinates, representing the cursor's position at each time step; 
\textbf{(ii)} timestamps; 
\textbf{(iii)} user identifiers; and 
\textbf{(iv)} session filenames. The dataset contains a total of $2,253,816$ samples. 

To utilize this dataset in the Q/H model, we binarize the data.
To elaborate, we first preprocess the movement data by calculating differences between successive cursor positions (\(\Delta x, \Delta y\)). Then, we calculate binary movement direction indicators using a binning strategy: 
if \(|\Delta x|\) is greater than or equal to \(|\Delta y|\), the movement is classified as horizontal (encoded as 1); otherwise, it is classified as vertical (encoded as 0). It should be noted that the DSSA algorithm is applicable to any discrete data, and the discretization process can be made more fine-grained by selecting larger alphabet sizes beyond binary.

\noindent\textbf{Touchalytics Dataset:} 
The Touchalytics dataset captures touch interaction patterns from $41$ users interacting with Android smartphones \cite{killourhy2009comparing}. The raw data includes the following components:
\textbf{(i)} phone ID, user ID, and document ID, \textbf{(ii)} time[ms], \textbf{(iii)} action, \textbf{(iv)} phone orientation, \textbf{(v)}x and y coordinates, \textbf{(vi)} pressure, area covered,  and finger orientation. The dataset contains a total of $912,132$ samples. We preprocess the raw data following the approach used for the Balabit dataset. That is, we encode movements into binary categories:
horizontal movements are encoded as 0, and vertical movements are encoded as 1.

\noindent\textbf{CMU Password Dataset}
The CMU Password dataset \cite{killourhy2009comparing} contains keystroke dynamics data collected from individuals typing a strong password. Each entry includes: 
\textbf{(i)} subject identifiers; 
\textbf{(ii)} session index; 
\textbf{(iii)} keystroke timings, such as hold time (H) and up-down time (UD), which record the duration and sequence of keystroke events. The dataset contains a total of $20,400$ samples. 
To analyze keystroke behavior, we preprocess the continuous timing data into binary sequences using the following steps:
\textbf{(i)} \textit{Normalization}: Keystroke timing data for each user is scaled between 0 and 1; 
\textbf{(ii)} \textit{Binary encoding}: Normalized values are converted into 3-bit binary sequences by discretizing them into eight bins with uniform thresholds, i.e. thresholds $\frac{i}{8},i\in [8]$; 
\textbf{(iii)} \textit{Time association}: Each keystroke event is linked to its respective binary sequence, repeating time values for consistency across the binary sequences.

\subsection{Model Description}

\noindent \textbf{MINE Architecture:}
We employ a neural network architecture consisting of an input layer with a single neuron, followed by two hidden layers, each with 100 neurons with Exponential Linear Unit (ELU) activation functions. The output layer has one neuron. We use an Adam optimizer with initial learning rate of $10^{-3}$, adjusted logarithmically every 30 epochs. Training is over 5,000 iterations per user pair and per input  sequence, with batch size of 100 samples.
For each user and binary sequence, we collect 8000 samples for each query pattern with length equal to two. In total we will have $8000\times m$ samples per  user training dataset.

\noindent\textbf{MLP Architecture:} The input layer has 1 neuron, followed by a hidden layer with 20 neurons, where batch normalization is applied, and with ReLU activation. The output layer contains 2 neurons, each corresponding to one of Bob's belief about users. The model is trained using the Adam optimizer with a learning rate of $10^{-3}$  and cross-entropy loss function. Training is conducted over 10 epochs for each user pair and sequence, with a batch size of 100 samples.

\noindent \textbf{DSSA Parameters:} We continue the test until all observations in the test data are exhausted, i.e., 
$t_{max}=\infty$ or Bob's belief falls outside of the interval $[0,01,0.99]$, i.e., $\epsilon_t=0.01$. 

\noindent\textbf{Simulation Hardware:} The simulations were run on a single NVIDIA A100-PCIE-40GB GPU.

\begin{figure*}[!h]
    \centering
    \includegraphics[width=\linewidth]{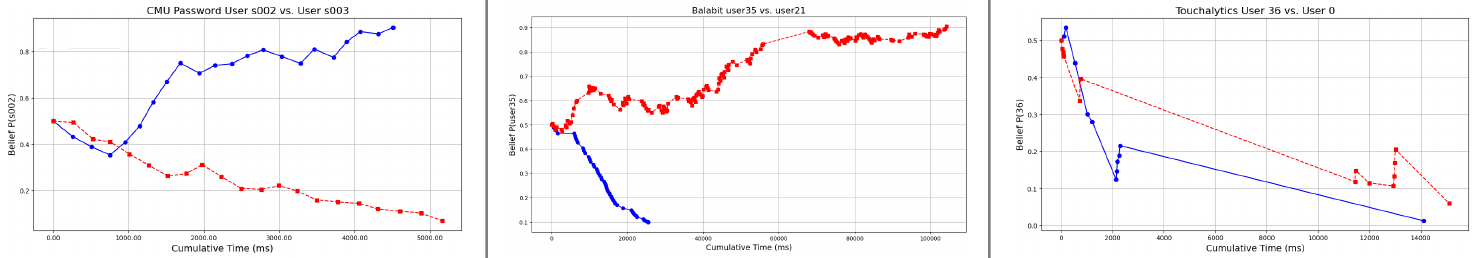}
    \caption{The trajectory of Bob's belief in the Random choice and DSSA algorithms for a single run in the CMU, Balabit, and Touchalytics datasets.}
    \label{fig:single_sim_real}
\end{figure*}

\begin{figure*}[!h]
    \centering
    \includegraphics[width=\linewidth]{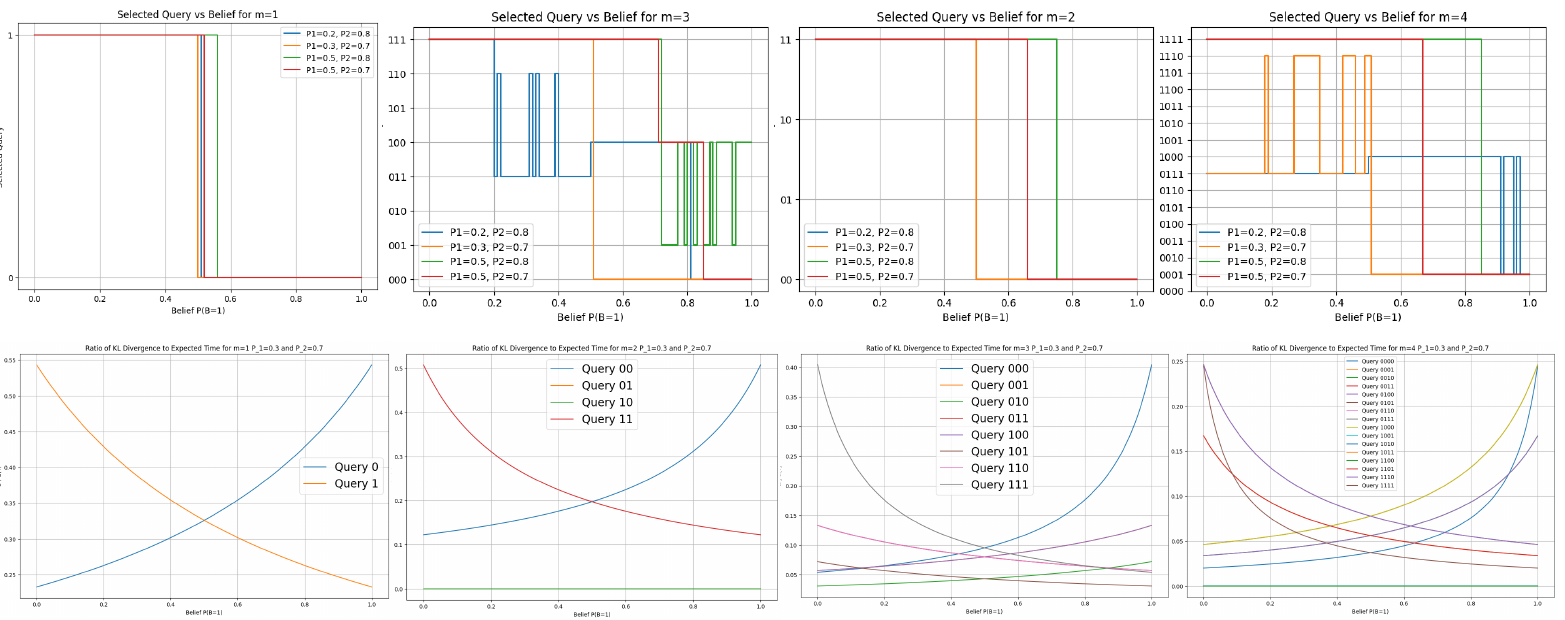}
    \caption{(Top) The optimal query choice in the IID setting under various probability distributions as function of Bob's belief. (Buttom) The change in the efficiency ratio of each query as a function of Bob's belief.}
    \label{fig:query_choice_IID}
\end{figure*}
\subsection{Additional Simulations in the Binary IID Setting}
\begin{figure*}[!h]
    \centering
    \includegraphics[width=\linewidth]{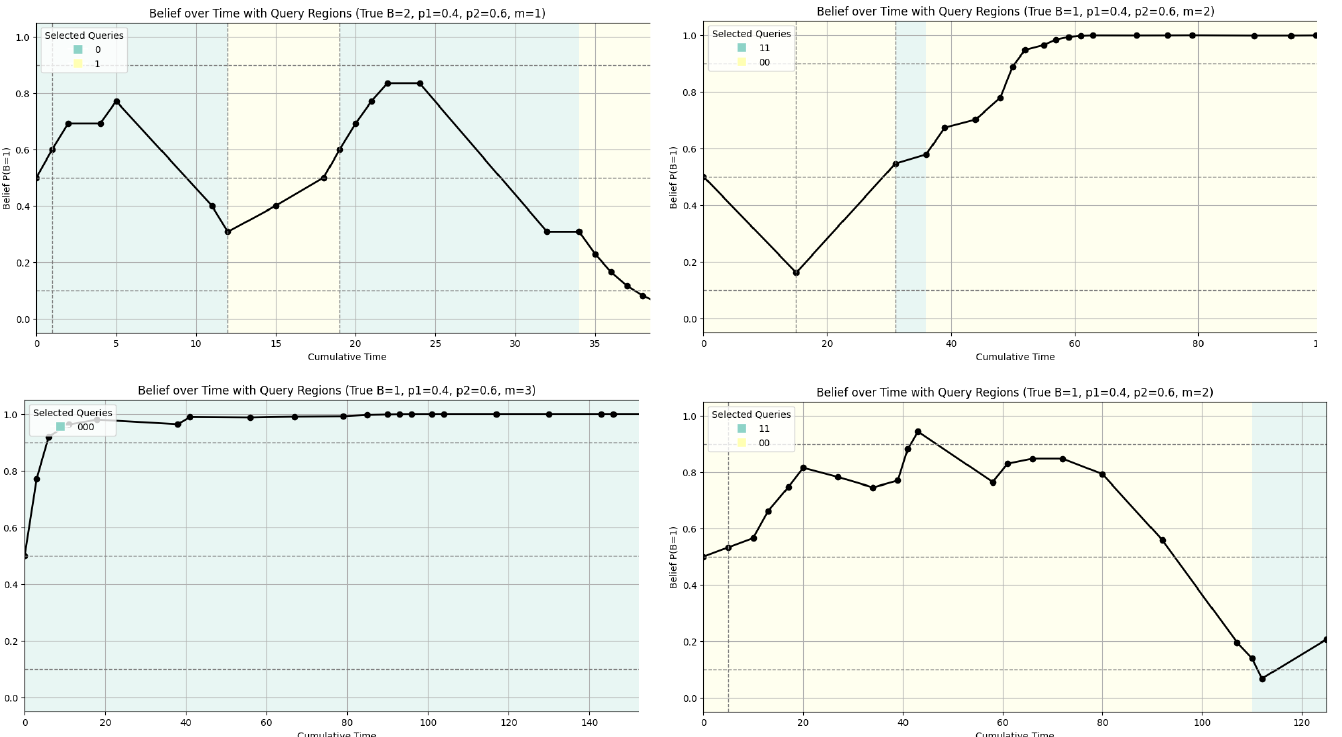}
    \caption{The belief trajectoty and choice of optimal query for $P(Z_1=1)=1-P(Z_2=1)=0.4$.}
    \label{fig:single_IID}
\end{figure*}

\noindent \textbf{Choice of Query As a Function of Bob's Belief.} In Theorem \ref{th:1}, we characterize the optimal static query function for the IID source setting. In particular, we showed that the choice depends on the KL-divergence between the hit time distributions of each query pattern under the two hypothesis, and the inverse of the expected hit time. The KL-divergence $D_{KL}(P_1(\Delta T)\parallel P_2(\Delta T)|\mathbf{q})$ does not depend on Bob's belief and hence is constant throughout the hypothesis testing process. However, the expected time $\mathbb{E}(\Delta T|\mathbf{q})$ is linearly related to Bob's belief $P(\widehat{\Omega}=1)$. The top plots in Figure \ref{fig:query_choice_IID} show the optimal query, characterized by Theorem \ref{th:1}, as a function of Bob's belief for various source distributions. We have taken $P(Z_1=1)< P(Z_2=1)$ in all cases. As expected, when Bob's belief $P(\Omega=1)$ is small (left side of each graph), the queries with more frequent ones are chosen, e.g., the all ones sequence, whereas if $P(\Omega=1)$ is large (right side of each graph), then, queries with more frequent zeros are chosen. The reason is that such query patterns are more likely to appear and the query response would be given in a shorter time on average. It can be observed that the optimal query sometimes fluctuates between two different patterns as a function of Bob's belief. For instance, for $m=4$ (the right-most figure), when Bob's belief changes between 0.2 and 0.5, and for $P(Z_1=1)=0.3, P(Z_2=1)=0.7$, the optimal query fluctuates between `1110' and `0111'. The reason for this observation is that the two queries have the exact same KL-divergence and expected hitting times, and the fluctuation is due to computation errors. To gain a deeper understanding about the choice of optimal query, the bottom four figures plot the efficiency ratio for all possible queries for $m=1,2,3,4$ and $P(Z_1=1)=0.3, P(Z_2=1)=0.7$. For instance, when $m=4$, the optimal query is `1110' (or `0111' alternatively) for $P(\Omega=1)<\frac{1}{2}$ and it is `0001' (or `1000', alternatively) for $P(\Omega=1)>\frac{1}{2}$.
\\\noindent\textbf{Evolution of Belief and Query Choice:} In Figure \ref{fig:single_IID}, we have shown the trajectory of Bob's belief as a function of number of queries in a single run of the hypothesis test for two IID binary sources with $P(Z_1=1)=0.4$ and $P(Z_2=1)=0.6$. It can be observed that the optimal query choice changes as a function of Bob's belief, and for belief values higher than $\frac{1}{2}$ (more likely $P(Z_1=1)=0.4)$, Bob chooses queries with more zero bit values, since they are more frequent, and for belief values lower than  $\frac{1}{2}$ (more likely $P(Z_2=1)=0.6)$, he chooses queries with more ones. 

\begin{figure*}[!t]
    \centering
    \includegraphics[width=0.85\linewidth]{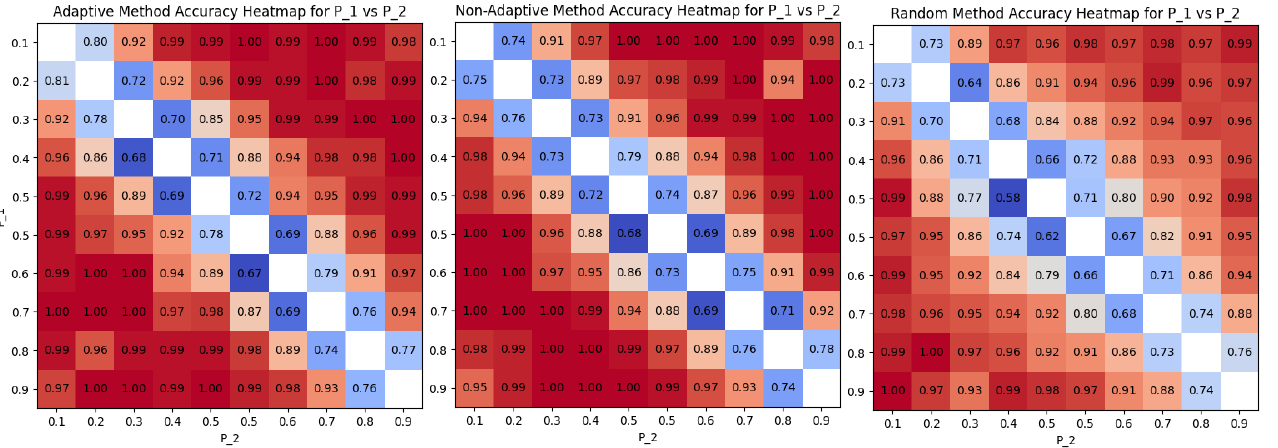}
    \caption{Accuracy heatmaps:  heuristic adaptive, optimal static, and Random Choice algorithms in IID settings. Alice's observations are restricted to 50 samples.}
    \label{fig:IID_Heatmap_2}
\end{figure*}

\noindent \textbf{Effects of Observation Length on Accuracy:} 
In Figure \ref{fig:IID_Heatmap}, we have provided the accuracy heatmap for the static, adaptive, and Random Choice algorithms. We have limited the length of the sequence observed by Alice to 20 bits, and the number of queries sent by Bob to 10. An interesting observation is that the top-right and bottom-left values, corresponding to $P_i\in \{0.1,0.9\}$, have low accuracies. The justification is that in these scenarios, although the sources are highly distinguishable, the algorithm may be \textit{unlucky} at initialization, and may choose the wrong pattern, e.g. if probability of $0$ is $0.1$ and the algorithm initiates with a a pattern with at least one 0, then it may have to wait a significant amount of time until the pattern is observed. This effect is especially significant in the simulation of Figure \ref{fig:IID_Heatmap} since Alice's observation is limited to 20 samples. To test this insight, we have simulated the same scenario with Alice's observation limited to 50 samples in Figure \ref{fig:IID_Heatmap_2}. We observe that our hypothesis is indeed true, and the accuracy of the aforementioned extreme cases improves with the increased sample size. It should be noted that as the observation length increases, Random Choice's performance becomes comparable to that of our proposed algorithms. In fact, for large observation lengths Random Choice may yield slightly better results, since the objective of our proposed methods is to optimize the tradeoff between time-to-detection and accuracy, and hence they may not be optimal if time-to-detection is not a factor. 

\subsection{Additional Simulations in the Real-World Settings}
\noindent \textbf{Evolution of Belief and Query Choice in Real-world Datasets:} In Figure \ref{fig:single_sim_real} we have plotted Bob's belief trajectory in a single run of the DSSA algorithm (blue curve) and the Random Choice Algorithm (red curve) when applied to the Balabit, Touchalytics, and CMU datasets. It can be observed that generally, the DSSA algorithm converges more quickly in all three datasets, and in the case of the Balabit dataset, it converges to the correct hypothesis whereas the Random Choice algorithm converges to the incorrect hypothesis. An interesting observation is that in all three real-world datasets, we observed a dominating query pattern for each user which consecutively optimized the efficiency ratio throughout the hypothesis test. Thus, the adaptive algorithm and the static algorithm have the exact same performance in these three runs. The interpretation is that the users have \textit{signature moves} which yield query patterns that dominate others in terms of information efficiency. We note that in the Touchalytics run, both the Random Choice and DSSA algorithms choose an inefficient query in one step, leading to a long wait time before a hit is observed. This is line with our other simulation in Figure \ref{fig:heatmap_datasets}, which shows that the proposed algorithms have better accuracy performance in the Balabit and CMU datasets, however, their accuracy performance in Touchalytics dataset is close to random guessing, hence, they may not choose efficient queries in general.

\end{document}